\documentclass{ws-procs975x65}

\def\beq{\begin{equation}}
\def\eeq{\end{equation}}

\begin{document}

\title{Gamma-ray burst cosmology: Hubble diagram and star formation history
\footnote{Based  on presentations  at  the  Fourteenth  Marcel  Grossmann  
Meeting  on  General  Relativity, Rome, July 2015.}}

\author{Jun-Jie Wei$^1$ and Xue-Feng Wu$^{1,2,\dagger}$}

\address{$^1$Purple Mountain Observatory, Chinese Academy of Sciences,\\
Nanjing 210008, China\\
$^2$Joint Center for Particle, Nuclear Physics and Cosmology,\\
Nanjing University-Purple Mountain Observatory, Nanjing 210008, China\\
$^\dagger$E-mail: jjwei@pmo.ac.cn; xfwu@pmo.ac.cn}

\begin{abstract}
We briefly introduce the disadvantages for Type Ia supernovae (SNe Ia) as
standard candles to measure the Universe, and suggest Gamma-ray bursts (GRBs)
can serve as a powerful tool for probing the properties of high redshift Universe.
We use GRBs as distance indicators in constructing the Hubble diagram
at redshifts beyond the current reach of SNe Ia observations.
Since the progenitors of long GRBs are confirmed to be massive stars,
they are deemed as an effective approach to study the cosmic star formation rate (SFR).
A detailed representation of how to measure high-$z$ SFR
using GRBs is presented. Moreover, first stars can form only in structures that
are suitably dense, which can be parameterized by defining the minimum
dark matter halo mass $M_{\rm min}$. $M_{\rm min}$ must play a crucial role
in star formation. The association of long GRBs with the collapses of
massive stars also indicates that the GRB data can be applied to constrain
the minimum halo mass $M_{\rm min}$ and to investigate star formation in dark matter halos.
\end{abstract}

\keywords{Gamma-ray bursts; standard candles; stars formation.}

\bodymatter


\section{Introduction}
Gamma-ray bursts (GRBs) are the most powerful explosions in the cosmos, which can be divided
into long GRBs with duration times $T_{90}>2$ s and short GRBs with $T_{90}<2$ s.\cite{Kouveliotou et al.1993}
In theory, it is generally accepted that long GRBs are formed by the core collapses of
massive stars,\cite{Woosley1993,Woosley et al.2006} while short GRBs are arised from the mergers of binary compact stars.\cite{Eichler et al.1989,Narayan1992}
Because of their high luminosities, GRBs can be discovered out to
very high redshifts. To date, the farthest burst detected is at $z=8.2$ (GRB 090423\cite{Tanvir et al.2009})
\footnote{A photometric redshift of $z\sim9.4$ for GRB 090429B was measured by Ref.~\refcite{Cucchiara et al.2011}.}.
GRBs are therefore considered as a powerful tool for probing the properties of the early Universe.

Type Ia supernovae (SNe Ia) have been treated as an ideal standard candle to measure dark energy
and cosmic expansion.\cite{Perlmutter et al.1998,Schmidt et al.1998,Riess et al.1998} However, the highest redshift of SNe Ia
so far is $z=1.914$.\cite{Jones et al.2013} It is hard to observe SNe Ia at redshifts $>2$, even with the excellent
space-based platforms such as SNAP.\cite{Sholl et al.2004} Since lots of the interesting evolution of the Universe
occurred before this epoch, the usage of SNe Ia in cosmology is limiting. In contrast to SNe Ia,
GRBs can be detected at higher redshifts. Moreover, gamma-ray photons from GRBs, unlike that
the optical photons of supernovae, are immune to dust extinction. The observed gamma-ray
flux is an actual measurement of the prompt emission flux. Thus, GRBs are potentially
a more promising cosmological probe than SNe Ia at higher redshifts. The possible use of GRBs
as ``relative standard candles'' started to become reality after some luminosity relations
between energetics and spectral properties were found.\cite{Amati et al.2002,Ghirlanda et al.2004,Yonetoku et al.2004,Liang et al.2005}
These GRB luminosity relations have been deemed as distance indicators
for cosmology. For example, Ref.~\refcite{Schaefer2003} constructed the first GRB Hubble diagram using two GRB luminosity indicators.
With the correlation between the collimation-corrected gamma-ray energy $E_{\rm \gamma}$ and
the spectral peak energy $E_{\rm p}$, Ref.~\refcite{Dai et al.2004} obtained tight constraints on cosmological parameters and
dark energy.  Ref.~\refcite{Ghirlanda et al.2004b} placed much tighter limits on cosmological parameters with
the same $E_{\rm p}-E_{\rm \gamma}$ relation and SNe Ia. Ref.~\refcite{Liang et al.2005} constrained cosmological
parameters and the transition redshift using a model-independent multivariable GRB luminosity relation.
Ref.~\refcite{Schaefer2007} constructed a GRB Hubble diagram with 69 bursts with the help of five luminosity indicators.
Ref.~\refcite{Kodama et al.2008} found that the time variation of the dark energy is very small or zero up to $z\sim6$ using
the relation between the isotropic peak luminosity $L$ and $E_{\rm p}$. Ref.\refcite{Tsutsui et al.2009}
extended the Hubble diagram up to $z=5.6$ based on 63 bursts using the $E_{\rm p}-L$ relation
and shown that these GRBs were in agreement with the concordance model within $2\sigma$ level.
In the meantime, a lot of works\cite{Firmani et al.2005,Xu et al.2005,Amati2006,Amati et al.2008,Liang et al.2006,Wang et al.2006,Liang et al.2008,Qi et al.2008a,Qi et al.2008b,Wei et al.2009,Yu et al.2009,Wei2010,Wang et al.2011a,Wei et al.2013,Ding et al.2015,Lin et al.2015,Izzo et al.2015,Wei et al.2015} have been done in this so-called GRB cosmology field.
Please see Refs.~\refcite{Ghirlanda et al.2006,Amati et al.2013,Wang et al.2015} for recent reviews.

With the improving observational techniques and a wider coverage in redshift,
we now have a better understanding of the star formation history in the Universe.
However, direct star formation rate (SFR) measurements are still quite difficult at
high redshifts ($z\geq6$), particularly at the faint end of the galaxy luminosity function.
Fortunately, the collapsar model suggests that long GRBs provide a complementary tool
for measuring the high-$z$ SFR from a different perspective, i.e., by the means of
investigating the death rate of massive stars rather than observing them directly during their lives.
Due to the fact that long GRBs are a product of the collapses of massive stars,
the cosmic GRB rate should in principle trace the cosmic SFR.
However, the Swift observations reveal that the GRB rate does not
strictly follow the SFR, but instead implying some kind of additional evolution.
\cite{Daigne et al.2006a,Guetta et al.2007,Le et al.2007,Salvaterra et al.2007,Kistler et al.2008,Kistler et al.2009,Salvaterra et al.2009,Campisi et al.2010,Qin et al.2010,Wanderman et al.2010,Cao et al.2011,Virgili et al.2011,Lu et al.2012,Robertson et al.2012,Salvaterra et al.2012,Wang2013,Wei et al.2014}
An enhanced evolution parametrized as $(1+z)^{\alpha}$ is usually adopted to describe
the difference between the GRB rate and the SFR.\cite{Kistler et al.2008}
In order to explain the observed discrepancy, several possible mechanisms have been proposed,
including cosmic metallicity evolution,\cite{Langer et al.2006,Li2008} stellar initial mass function evolution,\cite{Xu et al.2009,Wang et al.2011}
and luminosity function evolution.\cite{Virgili et al.2011,Salvaterra et al.2012,Tan et al.2013,Tan et al.2015}
Of course, if we knew the mechanism responsible for the discrepancy between
the GRB rate and the SFR, we could set a severe limit on the high-\emph{z} SFR using the GRB data alone.

In the framework of hierarchical structure formation, a self-consistent SFR model can be calculated.
Especially, the baryon accretion rate accounts for the structure formation process, which governs
the size of the reservoir of baryons available for star formation in dark matter halos, can be obtained
from the hierarchical scenario.
First stars can form only in structures that are suitably dense, which can be parametrized
by defining the minimum mass $M_{\rm min}$ of a dark matter halo of the collapsed structures where
star formation occurs. In briefly, structures with masses smaller than $M_{\rm min}$ are served as part of
the intergalactic medium and do not participate in the process of star formation. The minimum halo mass
$M_{\rm min}$ must, therefore, play a crucial role in star formation. There are a few direct constraints on
$M_{\rm min}$ as follows. In order to simultaneously reproduce the current baryon fraction and
the early chemical enrichment of the intergalactic medium, Ref.~\refcite{Daigne et al.2006b} suggested that
a minimum halo mass of $10^{7}-10^{8}M_{\odot}$ and a moderate outflow efficiency were required.
With a minimum halo mass of $M_{\rm min}\simeq10^{11}M_{\odot}$, Ref.~\refcite{Bouche et al.2010} could
explain both the observed slopes of the star formation rate--mass and Tully--Fisher relations.
Ref.~\refcite{Munoz et al.2011} found that the minimum halo mass can be constrained by matching the observed
galaxy luminosity distribution, in which $M_{\rm min}$ was limited to be
$10^{8.5}\rm M_{\odot}<M_{\rm min}<10^{9.7}\rm M_{\odot}$ at the $95\%$ confidence level.
The association of long GRBs with the death of massive stars provides a new interesting tool
to investigate star formation in dark matter halos.

In this work, we review the applications of GRBs in cosmology.
The rest of this paper is arranged as follows. In Section~2,
we construct the GRB Hubble diagram and describe its cosmological implications.
The constraints on the high-$z$ SFR from GRBs are presented in Section~3,
and the capability of GRBs to probe star formation in dark matter halos is presented in Section~4.
Finally, the conclusions and discussion are drawn in Section~5.
For the more details of full samples and analysis results we discussed here, please refer to Refs.~\refcite{Wei et al.2013,Wei et al.2014,Wei et al.2016}.

\section{The GRB Hubble Diagram and Its Cosmological Implications}

\subsection{The latest luminosity relation}
The isotropic equivalent gamma-ray energy $E_{\gamma,{\rm iso}}$ of GRBs can be calculated with
\begin{equation}
E_{\gamma,{\rm iso}}=\frac{4\pi D^{2}_{L}(z)S_\gamma}{(1+z)}K,
\end{equation}
where $D_{L}(z)$ represents the luminosity distance at redshift \emph{z},
$S_\gamma$ is the observed gamma-ray fluence, and $K$ is a factor used to
correct the observed fluence within an observed bandpass to a broad band
(i.e., $1-10^{4}$ keV in this paper) in the rest frame.
In the standard $\Lambda$CDM model, the luminosity distance is given as
\begin{equation}
D_{L}(z)={c\over H_{0}}{(1+z)\over\sqrt{\mid\Omega_{k}\mid}}\; {\rm sinn}\left\{\mid\Omega_{k}\mid^{1/2}
\times\int_{0}^{z}{dz\over\sqrt{\Omega_{m}(1+z)^{3}+\Omega_{k}(1+z)^{2}+\Omega_{\Lambda}}}\right\}\;.
\end{equation}
Here, sinn is $\sinh$ when the spatial curvature $\Omega_{k}>0$ and
$\sin$ when $\Omega_{k}<0$. For a flat Universe with $\Omega_{k}=0$,
Equation~(2) reduces to the form $(1+z)c/H_{0}$ times the integral.

Many efforts have been made to seek other distance indicators from
the GRB spectra and light-curves that provide more precise constraints on the luminosity.
Inspired by the work of Ref.~\refcite{Liang et al.2005}, we search for the empirical luminosity relation
between $E_{\gamma,{\rm iso}}$, $E'_{\rm p}$, and $t'_{\rm b}$, which is known as
the Liang-Zhang relation. We collect a sample of 33 high-quality GRBs, each burst has a measurement of \emph{z},
the spectral peak energy $E_{\rm p}$, and the jet break time $t_{\rm b}$ observed
in the optical band. The form of this luminosity correlation can be expressed as:
\begin{equation}
\log E_{\gamma,{\rm iso}}=\kappa_{0}+\kappa_{1}\log E'_{\rm p}+\kappa_{2}\log t'_{\rm b}\;,
\end{equation}
where $E'_{\rm p}=E_{\rm p}(1+z)$ in keV and $t'_{\rm b}=t_{\rm b}/(1+z)$ in days.
To find the best-fit coefficients $\kappa_0$, $\kappa_1$ and $\kappa_2$,
we adopt the method described in Ref.~\refcite{D'Agostini2005}. First, we simplify the
notation by writing $x_{1}=\log E'_{\rm p}$, $x_{2}=\log t'_{\rm b}$, and
$y=\log E_{\gamma,{\rm iso}}$. The joint likelihood function for the
coefficients $\kappa_{0}$, $\kappa_{1}$, $\kappa_{2}$ and the intrinsic
scatter $\sigma_{\rm int}$ thus can be written as
\begin{equation}
\begin{split}
L(\kappa_{0},\kappa_{1},\kappa_{2},\sigma_{\rm int})\propto\prod_{i}\frac{1}{\sqrt{\sigma^{2}_{\rm int}+\sigma^{2}_{y_{i}}+\kappa^{2}_{1}\sigma^{2}_{x_{1,i}}+\kappa^{2}_{2}\sigma^{2}_{x_{2,i}}}}\times\qquad\qquad\\
\null\qquad\exp\left[-\frac{(y_{i}-\kappa_{0}-\kappa_{1}x_{1,i}-\kappa_{2}x_{2,i})^{2}}{2(\sigma^{2}_{\rm int}+\sigma^{2}_{y_{i}}+\kappa^{2}_{1}\sigma^{2}_{x_{1,i}}+\kappa^{2}_{2}\sigma^{2}_{x_{2,i}})}\right]\;,
\end{split}
\end{equation}
where $i$ denotes the corresponding serial number of each burst in our sample.

\begin{figure}[h]
\vskip-0.1in
\begin{center}
\includegraphics[width=3.0in]{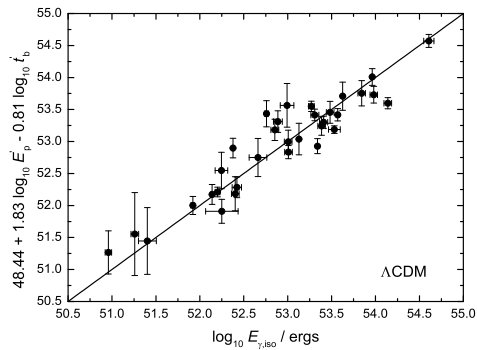}
\end{center}
\caption{The latest $E_{\gamma,{\rm iso}}$ versus $E'_{\rm p}-t'_{\rm b}$ correlation.
The solid line shows the best-fitting results.}
\label{fig1}
\end{figure}

Fig.~\ref{fig1} shows the best-fit luminosity correlation. In this calculation,
a flat $\Lambda$CDM cosmology with $\Omega_{m}=0.29$ and $H_{0}=69.32$ km s$^{-1}$ Mpc$^{-1}$
from the 9-yr WMAP data is adopted.\cite{Bennett et al.2013} Using the
above methodology, we find that the best-fit correlation between
$E_{\gamma,{\rm iso}}$ and $E'_{\rm p}-t'_{\rm b}$ is
\begin{equation}
\log E_{\gamma,{\rm iso}}=(48.44\pm0.38)+(1.83\pm0.15)\log E'_{\rm p}
-(0.81\pm0.22)\log t'_{\rm b}\;,
\end{equation}
with an intrinsic scatter $\sigma_{\rm int}=0.25\pm0.06$.
The best-fitting line is also plotted in Fig.~\ref{fig1}.

\subsection{Constraints on cosmological parameters}
\begin{figure}[h]
\begin{center}
\includegraphics[width=2.9in]{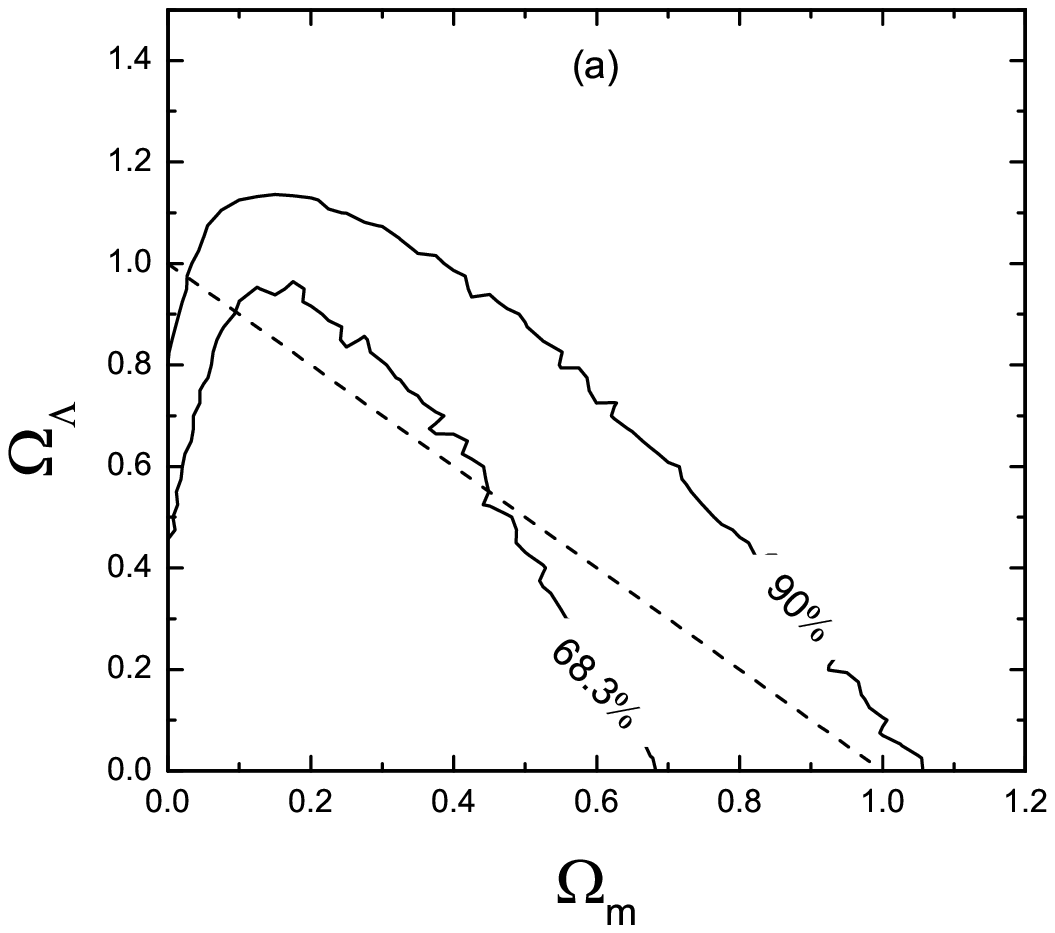}\hskip-0.8in
\includegraphics[width=2.9in]{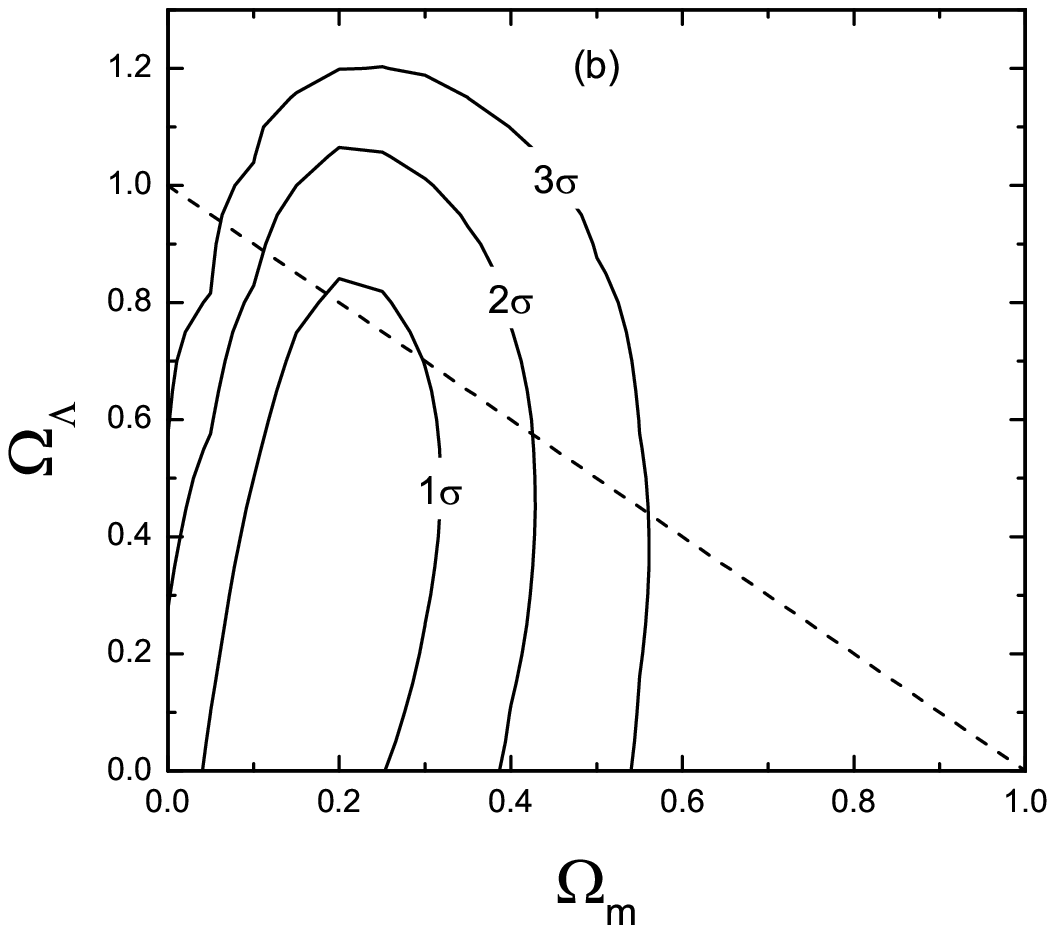}
\end{center}
\caption{(a) Contour confidence levels of $\Omega_{m}$ and $\Omega_{\Lambda}$ from the GRB data,
using method I. (b) Same as panel (a), except now using method II.}
\label{fig2}
\end{figure}

The dispersion of the Liang-Zhang correlation is so small that it
has been deemed as a good luminosity indicator for cosmology.\cite{Liang et al.2005,Wang et al.2006}
However, this correlation is cosmology-dependent, i.e.,
$E_{\gamma,{\rm iso}}$ is a function of cosmological parameters,
we cannot therefore use it to constrain the cosmological parameters directly. In order to
avoid circularity issues, we use the following two methods to overcome
this problem:

\emph{Method I}. We repeat the above analysis while varying the cosmological
parameters $\Omega_{m}$ and $\Omega_{\Lambda}$.\cite{Amati et al.2008,Ghirlanda2009}
In other words, the coefficients of this luminosity correlation are optimized simultaneously
with the cosmological parameters now. The contours in Fig.~\ref{fig2}(a) show that $\Omega_{m}$ and
$\Omega_{\Lambda}$ are weakly limited; only an upper limit of $\sim0.68$ and $\sim0.95$
can be obtained at $1\sigma$ for $\Omega_{m}$ and $\Omega_{\Lambda}$. But,
if we just consider a flat universe (dashed line), the constraints on the cosmological parameters
can be tightened at the $1\sigma$ level, for which $0.10<\Omega_{m}<0.45$ and $0.55<\Omega_{\Lambda}<0.90$.
The most likely values of $\Omega_{m}$ and $\Omega_{\Lambda}$ are $(0.22_{-0.12}^{+0.23}, 0.78_{-0.23}^{+0.12})$.

\emph{Method II}. We also use the approach of Ref.~\refcite{Liang et al.2005} to circumvent the circularity problem,
which is based on the calculation of the probability function for a given set of cosmological parameters.
We refer the reader to Ref.~\refcite{Liang et al.2005} for more details. With this method, the $1\sigma$ -- $3\sigma$
constraint contours of the probability in the ($\Omega_{m}$, $\Omega_{\Lambda}$) plane are shown in Fig.~\ref{fig2}(b).
These contours show that at the $1\sigma$ level, $0.04<\Omega_{m}<0.32$, while $\Omega_{\Lambda}$ is poorly
limited; only an upper limit of $\sim0.84$ can be obtained at this confidence level. However, if we just consider
a flat Universe (dashed line), the allowed region at the $1\sigma$ level is restricted to be $0.19<\Omega_{m}<0.30$
and $0.7<\Omega_{\Lambda}<0.81$. The best-fit values correspond to $(\Omega_{m},\Omega_{\Lambda})=
(0.25_{-0.06}^{+0.05}, 0.75_{-0.05}^{+0.06})$.

\begin{figure}[h]
\vskip-0.1in
\begin{center}
\includegraphics[width=3.5in]{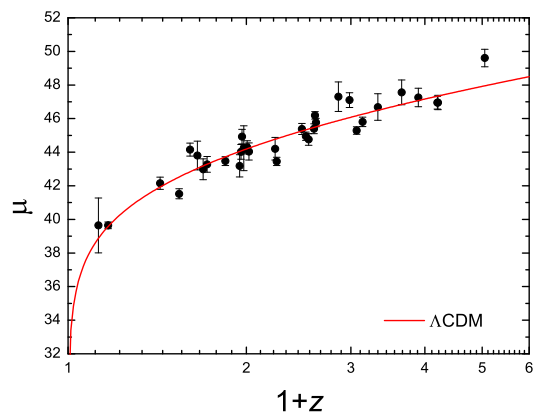}
\end{center}
\caption{Hubble diagram for 33 GRBs. The solid line shows the theoretical $\mu$.}
\vskip-0.2in
\label{fig3}
\end{figure}

With the best-fit flat $\Lambda$CDM model ($\Omega_{m}=0.25$ and $\Omega_{\Lambda}=0.75$),
the theoretical distance modulus $\mu_{\rm th}$ is estimated as
\begin{equation}
\mu_{\rm th}(z)\equiv5 \log\left(D_{L}(z)/10 \;\rm pc\right)\;.
\end{equation}
While, the observed distance modulus $\mu_{\rm obs}$ of each GRB can be calculated using the best-fit Liang-Zhang relation,
i.e.,
\begin{equation}
\mu_{\rm obs}=2.5[\kappa_{0}+\kappa_{1} \log E'_{\rm p}+\kappa_{2} \log t'_{\rm b}-\log(4\pi S_{\gamma}K)
+\log(1+z)]-97.45\;.
\end{equation}
The Hubble diagram of GRBs then is constructed in Fig.~\ref{fig3}, together with the best-fit theoretical line.

\section{Measuring the High-\emph{z} Star Formation Rate}

The connection of long gamma-ray bursts (LGRBs) with the core collapse of massive
stars has been strongly confirmed by many associations between LGRBs and Type Ic
supernovae,\cite{Hjorth et al.2003,Stanek et al.2003} which may provides a good opportunity
for constraining the high-\emph{z} SFR.\cite{Chary et al.2007,Yuksel et al.2008,Trenti et al.2012}
Owing to the short lives of massive stars, the SFR can be regarded as their death rate approximatively.

\begin{figure}[h]
\vskip-0.1in
\begin{center}
\includegraphics[width=3.5in]{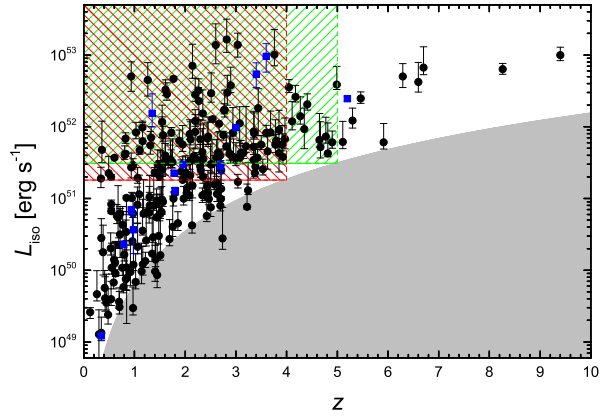}
\end{center}
\caption{The luminosity-redshift distribution of 244 \emph{Swift} LGRBs.
The blue squares are 13 dark bursts with firm redshift measurements.
The gray shaded region represents below the \emph{Swift} detection threshold.
The boxes represent 118 GRBs with $z<4$ and $L_{\rm iso}>1.8\times10^{51}$ erg $\rm s^{-1}$ (red),
and 104 GRBs with $z<5$ and $L_{\rm iso}>3.1\times10^{51}$ erg $\rm s^{-1}$ (green), respectively.}
\vskip-0.1in
\label{fig4}
\end{figure}

Fig.~\ref{fig4} shows the luminosity-redshift distribution of 244 \emph{Swift} GRBs, in which
the gray shaded region approximates the detection threshold of \emph{Swift}.
For more details on this GRB sample, please see Ref.~\refcite{Wei et al.2014}.
The luminosity threshold can be estimated using a bolometric flux limit
$F_{\rm lim}=1.2\times10^{-8}$ erg $\rm cm^{-2}$ $\rm s^{-1}$,\cite{Kistler et al.2008}
i.e., $L_{\rm lim}=4\pi D_{L}^{2}F_{\rm lim}$. The \emph{Swift}/Burst Alert Telescope (BAT)
trigger is so complex that it is hard to parametrize exactly the sensitivity of BAT.\cite{Band2006}
Furthermore, although the SFR density is well measured at relatively low redshifts
($z\leq4$), it is poorly constrained at $z\geq4$. To avoid the complications that
would be caused by the use of a detailed treatment of the \emph{Swift} threshold
and the SFR at high-\emph{z}, we employ a model-independent approach by selecting
only bursts with $L_{\rm iso}>L_{\rm lim}$ and $z<4$, as Ref.~\refcite{Kistler et al.2008} did in their treatment.
The cut in luminosity is chosen to be equal to the threshold at the maximum redshift
of the sample, i.e., $L_{\rm lim}(z=4)\approx1.8\times10^{51}$ erg $\rm s^{-1}$.
The cut in luminosity and redshift can reduce the selection effects by removing lots of low-$z$, low-$L_{\rm iso}$
bursts that could not be detected at higher redshift. With these conditions, our final
tally of GRBs is 118. These data are delimited by the red box in Fig.~\ref{fig4}.

Since the use of luminosity cuts, the integral of the luminosity function
can be treated as a constant coefficient, no matter what the specific form of
the luminosity function happens to be. Assuming that the relationship between
the comoving GRB rate and the SFR density $\dot{\rho}_{\star}$ can be described as
$\dot{n}_{\rm GRB}(z)=\varepsilon(z)\dot{\rho}_{\star}(z)$, where $\varepsilon(z)$
represents the formation efficiency of LGRBs. Now, the expect number of GRBs within
a redshift range $(0,\;z)$ can be written as
\begin{equation}\label{SMALL}
\begin{split}
N(<z)\propto \int_{0}^{z}\dot{\rho}_{\star}(z)\frac{\varepsilon(z)}{1+z}\frac{{\rm d} V_{\rm com}}{{\rm d} z}\;{\rm d}z\;,
\end{split}
\end{equation}
where $\rm d \emph{V}_{com}/d\emph{z}$ is the comoving volume element, and
the factor $(1+z)^{-1}$ accounts for the cosmological time dilation.
For relatively low redshifts ($z \leq 4$), the SFR density $\dot{\rho}_{\star}$
has been well fitted with a piecewise power law,\cite{Li2008,Hopkins et al.2006}
for which
\begin{equation}\label{SFRone}
\log \dot{\rho}_{\star}(z)=a+b\log(1+z)\;,
\end{equation}
where
\begin{equation}
  (a,b) = \left\lbrace \begin{array}{ll}(-1.70, 3.30), ~~~~~~~~~~~z <0.993\\
                                           (-0.727, 0.0549), ~~~~~0.993< z <3.8, \\
\end{array} \right.
\end{equation}
and $\dot{\rho}_{\star}$ is in units of $\rm M_{\bigodot}$ $\rm yr^{-1}$ $\rm Mpc^{-3}$.

\begin{figure}[h]
\vskip-0.1in
\begin{center}
\includegraphics[width=2.5in]{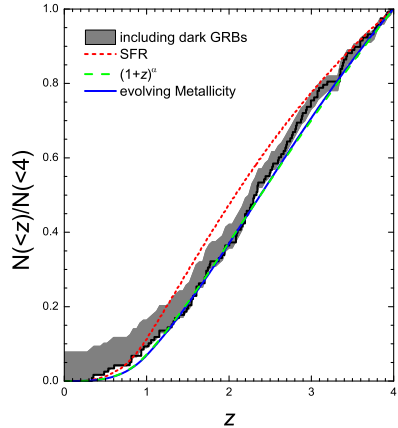}
\end{center}
\vskip-0.2in
\caption{Cumulative redshift distribution of 118 \emph{Swift} LGRBs
with $z<4$ and $L_{\rm iso}>1.8\times10^{51}$ erg $\rm s^{-1}$ (steps).
The red dotted line represents the GRB rate inferred from the SFR.
The green dashed line shows the GRB rate inferred from the SFR including $(1+z)^{0.8}$ evolution.
The blue solid line corresponds to the GRB rate inferred from the SFR including an evolving metallicity.}
\label{fig5}
\end{figure}

Fig.~\ref{fig5} shows the cumulative redshift distribution of 118 GRBs (steps), normalized over the redshift range
$0<z<4$. In Fig.~\ref{fig5}, we compare the observed redshift distribution with three kinds of redshift evolution,
characterized through the function $\varepsilon(z)$. If the non-evolution case (i.e., the function
$\varepsilon(z)$ is constant) is considered, we find that the expectation from the SFR alone (dotted red line)
does not provide a good representation of the observations. If we consider the density evolution model
(i.e., $\varepsilon(z)\propto(1+z)^{\alpha}$), we find that the $\chi^{2}$ statistic is minimized for
$\alpha=0.8$ and the weak redshift evolution ($\alpha=0.8$) can reproduce the observed redshift distribution
quite well (dashed green line). At the $2\sigma$ confidence level, the value of $\alpha$ is in the range $0.07<\alpha<1.53$.
It has been proposed that the observationally required evolution may be due to an evolving metallicity.\cite{Langer et al.2006,Li2008}
To test this interpretation of the anomalous evolution, we
assume that the GRB rate traces the SFR and a cosmic evolution in metallicity,
i.e., $\varepsilon(z) \propto \Theta(Z_{\rm th},z)$.
Here, $\Theta(Z_{\rm th},z)$ accounts for the fraction of galaxies at redshift $z$ with metallicity below $Z_{\rm th}$,\cite{Langer et al.2006}
which can be expressed as $\Theta(Z_{\rm th},z)=\hat{\Gamma}\left[\kappa+2,
(Z_{\rm th}/Z_{\odot})^{\beta} 10^{0.15 \beta z}\right]/\Gamma(\kappa+2)$, where $Z_{\odot}$ is the solar
metal abundance, $\hat{\Gamma}(a,x)$ and $\Gamma(x)$ are the incomplete and complete Gamma functions,
$\kappa=-1.16$ and $\beta=2$.\cite{Panter et al.2004,Savaglio2006} We find that the observations
can be well fitted if $Z_{\rm th}=0.52Z_{\odot}$ is adopted (blue line). At the $2\sigma$ confidence level,
the value of $Z_{\rm th}$ lies in the range $0.19Z_{\odot}<Z_{\rm th}<0.85Z_{\odot}$. A comparison between
this theoretical curve and that obtained with $\varepsilon(z)\propto(1+z)^{0.8}$ shows that
the differences between these two fits is very slight. Therefore,
we confirm that the anomalous evolution may be due to the cosmic metallicity evolution.

In the following, we will consider the implications of these findings for the high-$z$ SFR,
assuming that GRBs follow the star formation history and an additional evolution characterized by
$(1+z)^{\alpha}$. We will use the best fitting value $\alpha=0.8$ for a reasonable
description of this evolutionary effect.

\begin{figure}[h]
\vskip-0.1in
\begin{center}
\includegraphics[width=3.0in]{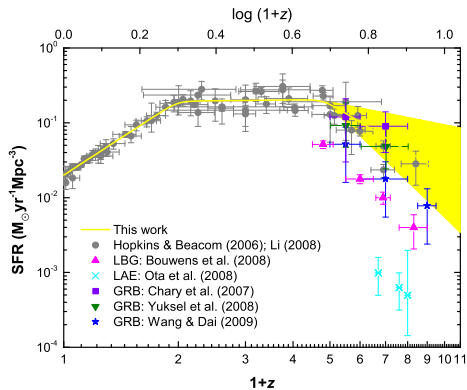}
\end{center}
\vskip-0.2in
\caption{The cosmic star formation history. The high-\emph{z} SFR
(orange shaded region) is constrained by the \emph{Swift} GRB data,
and is characterized by a power-law index $-5.07<\delta<-1.05$.
Some available SFR data are also shown for comparison.}
\label{fig6}
\end{figure}

The SFR is now well measured from $z=0$ to 4, but it is not well limited at high-\emph{z}
($z \geq 4$). In our analysis, a free parameter $\delta$ will be introduced to parametrize
the high-\emph{z} history as a power law at redshifts $z\geq 3.8$:
\begin{equation}\label{HSFR}
\dot{\rho}_{\star}(z) = 0.20\left(\frac{1+z}{4.8}\right)^{\delta},
\end{equation}
and this index $\delta$ will be constrained by the GRB observations.
The normalization constant in Equation~(11) is set by the requirement that
$\dot{\rho}_{\star}$ be continuous across $z=3.8$.
We optimize the index $\delta$ of high-\emph{z} SFR by minimizing the $\chi^{2}$ statistic
jointly fitting the observed redshift distribution and luminosity distribution of GRBs in our sample.
In the density evolution model, the best-fit is produced with a high-\emph{z} SFR with index
$\delta=-3.06_{-2.01}^{+2.01}(1\sigma)$. The range of high-\emph{z} star formation history with
$\delta\in(-5.07,-1.05)$ is marked with an orange shaded band in Fig.~\ref{fig6}, in comparison with
other observationally determined SFR data.\cite{Li2008,Chary et al.2007,Yuksel et al.2008,Hopkins et al.2006,Bouwens et al.2008,Ota et al.2008,Wang et al.2009}

Meanwhile, GRBs have indeed already been used to measure the high-$z$ SFR in several works.
For example, with deep observations of three $z\sim5$ GRBs from the \emph{Spitzer} Space Telescope,
a lower limit to the SFR at $z=4.5$ and $6$ was presented in Ref.~\refcite{Chary et al.2007}.
Ref.~\refcite{Yuksel et al.2008} subsequently made a new determination of the SFR at $z=4-7$ using the $Swift$
GRB data and found that no steep drop exists in the SFR up to $z\sim6.5$.
The use of four years of \emph{Swift} GRB observations as cosmological probes for the early Universe
was discussed by Ref.~\refcite{Kistler et al.2009}, who confirmed that the implied SFR to
$z \geq 8$ was consistent with Lyman Break Galaxy-based measurements.
Ref.~\refcite{Wang et al.2009} used a sample of 119 GRBs to constrain the high-\emph{z}
SFR up to $z\sim8.3$, showing that the SFR at $z > 4$ has a steep
decay with a slope of $\sim-5.0$. Based on the principal component analysis method,
Ref.~\refcite{Ishida et al.2011} probed cosmic star formation history up to $z \approx 9.4$ from the GRB data
and suggested that the level of star formation activity at $z \approx 9.4$ could
have been already as high as the present-day one ($\approx 0.01$
$M_{\bigodot}$ $\rm yr^{-1}$ $\rm Mpc^{-3}$).

\section{Probing Star Formation in Dark Matter Halos}
There is a general agreement about the fact that the rate of LGRBs does not strictly
trace the SFR but is actually enhanced by some unknown mechanisms at high-\emph{z}.
Many possible interpretations of the high-\emph{z} GRB rate excess have been introduced,
such as the GRB rate density evolution,\cite{Kistler et al.2008,Kistler et al.2009}
the cosmic metallicity evolution,\cite{Langer et al.2006,Li2008}
an evolving stellar initial mass function,\cite{Xu et al.2009,Wang et al.2011}
and an evolution in the break of luminosity function\cite{Virgili et al.2011,Salvaterra et al.2012,Tan et al.2013,Tan et al.2015}

Nevertheless, it should be underlined that the exception on the LGRB rate relates strongly to
the SFR models we adopted. With different star formation history models, the results on the difference
between the LGRB rate and the SFR could change on some level.\cite{Virgili et al.2011,Hao et al.2013}
Several forms of SFR are available in the literature. Most of previous works\cite{Kistler et al.2008,Kistler et al.2009,Li2008,Salvaterra et al.2009,Salvaterra et al.2012,Robertson et al.2012}
were focused on the the widely accepted SFR model of Ref.~\refcite{Hopkins et al.2006}, which provides a good piecewise-linear fit to
the numerous multiband observations. However, it is obviously that this empirical fit will
vary depending on the observational data and the functional form used. Ref.~\refcite{Hao et al.2013} confirmed that
the LGRB rate was still biased tracer of this empirical SFR model. On the contrast,
using the self-consistent SFR model calculated from the hierarchical structure formation scenario,
they found that a significant fraction of LGRBs occur in dark matter halos with mass down to $10^{8.5}M_{\odot}$ could give an
alternative explanation for the discrepancy between the LGRB rate and the SFR.

The fact that stars can form only in structures that are suitably dense, which can be defined
by the minimum mass $M_{\rm min}$ of a dark matter halo of the star-forming structures.
Thus, no stars will be generated in dark matter halos smaller than $M_{\rm min}$.
The minimum halo mass $M_{\rm min}$ must play a crucial role in star formation.
The collapsar model suggests that LGRBs are a new promising tool for probing star formation
in dark matter halos. In principle, the expected redshift distributions of LGRBs
can be calculated from the self-consistent CSFR model as a function of $M_{\rm min}$.
Therefore, the minimum halo mass $M_{\rm min}$ can be well constrained by comparing
the observed and expected redshift distributions.

\begin{figure}[h]
\vskip-0.1in
\begin{center}
\includegraphics[width=3.0in]{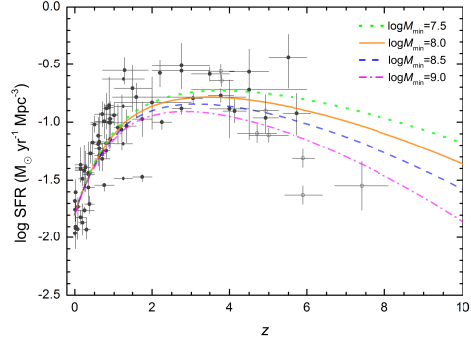}
\end{center}
\vskip-0.2in
\caption{The cosmic SFR as a function of redshift.
The curves correspond to the self-consistent models with
a minimum halo mass $M_{\rm min}=10^{7.5}$ $\rm M_{\odot}$, $10^{8.0}$ $\rm M_{\odot}$,
$10^{8.5}$ $\rm M_{\odot}$, and $10^{9.0}$ $\rm M_{\odot}$, respectively.
The observational data are taken from Refs.~89,90 (dots) and Ref.~62 (circles).}
\label{fig7}
\end{figure}

Here, we adopt the self-consistent SFR model of Ref.~\refcite{Pereira et al.2010}.
In the framework of hierarchical structure formation, Ref.~\refcite{Pereira et al.2010}
obtained the CSFR by solving the evolution equation of the total gas density that includes
the baryon accretion rate, the gas ejection by stars, and the formation of stars
through the transfer of baryons in the dark matter halos.
In Fig.~\ref{fig7}, we show the SFR derived from the self-consistent models with different
values of the minimum halo mass $M_{\rm min}$.
The observational SFR data taken from Refs.~\refcite{Li2008,Hopkins2004,Hopkins2007}.
are also shown for comparison. As can be seen, the SFR $\dot{\rho}_{\star}(z)$
is very sensitive to the numerical value of $M_{\rm min}$, especially at high-\emph{z}.
One can also see from this plot that all of these models are in good agreement with
observational data at $z\leq6$.

In order to compare the observed redshift distribution of LGRBs, we calculate the expected redshift
distribution as
\begin{equation}
\frac{\rm d\emph{N}}{\rm d\emph{z}}=F(z)\frac{\varepsilon(z)\dot{\rho}_{\star}(z)}{\langle f_{\rm beam} \rangle}
\frac{\rm d \emph{V}_{com}/d\emph{z}}{1+z} \;,
\end{equation}
where $\langle f_{\rm beam} \rangle$ denotes the beaming factor and $F(z)$ accounts for
the ability both to trigger the burst and to obtain the redshift. Ref.~\refcite{Kistler et al.2008} pointed out that
$F(z)$ can be treated as a constant coefficient ($F_{0}$) when we just consider the bright
GRBs with luminosities enough to be detected within an entire redshift range.
As discussed above, the physical nature of the observed enhancement in the LGRB rate remains
under debate. For simplicity, we parametrize the redshift evolution in the LGRB rate as
$\varepsilon(z)=\varepsilon_{0}(1+z)^{\alpha}$, where $\varepsilon_{0}$ represents a constant
that takes into account the fraction of stars that produce LGRBs.
Since the evolutionary index $\alpha$ is conservatively kept as a free parameter, we have
two free parameters $M_{\rm min}$ and $\alpha$ in our calculation.

The theoretically expected number of LGRBs within a redshift bin $z_{1}\leq z \leq z_{2}$, for each
combination $\textbf{P}\equiv\{M_{\rm min}, \alpha\}$, can be expressed as
\begin{equation}
\begin{split}
N(z_{1}, z_{2}; \textbf{P})=\Delta t \frac{\Delta \Omega}{4\pi}\int^{z_{2}}_{z_{1}}F(z)\varepsilon(z)
\frac{\dot{\rho}_{\star}(z; M_{\rm min})}{\langle f_{\rm beam} \rangle}\frac{\rm d \emph{V}_{com}/d\emph{z}}{1+z}\; \rm d\emph{z}\\
=\mathcal{A}\int^{\emph{z}_{2}}_{\emph{z}_{1}}(1+\emph{z})^{\alpha}\dot{\rho}_{\star}(\emph{z}; M_{\rm min})\frac{\rm d \emph{V}_{com}/d\emph{z}}{1+\emph{z}}\; \rm d\emph{z}\;,
\end{split}
\end{equation}
where $\mathcal{A}=\Delta t \Delta \Omega F_{0}\varepsilon_{0}/4\pi\langle f_{\rm beam} \rangle$
is a constant that depends on the observed time, $\Delta t$, and the sky coverage, $\Delta \Omega$.
The constant $\mathcal{A}$ can be removed by normalizing the cumulative redshift distribution of GRBs
to $N(0, z_{\rm max})$, as
\begin{equation}
N(<z|z_{\rm max})=\frac{N(0, z)}{N(0, z_{\rm max})} \;.
\end{equation}

\begin{figure}[h]
\vskip-0.1in
\begin{center}
\includegraphics[width=5.0in]{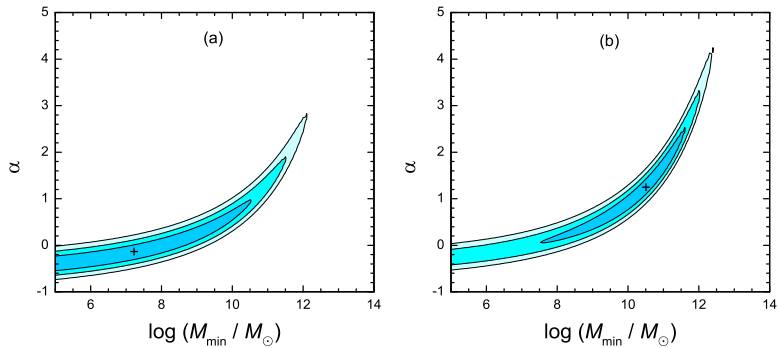}
\end{center}
\vskip-0.1in
\caption{(a): Contour confidence levels of $M_{\rm min}$ and $\alpha$,
    inferred from the redshift distribution of 118 \emph{Swift} GRBs with $z<4$ and
    $L_{\rm iso}>1.8\times10^{51}$ erg $\rm s^{-1}$. The cross represents the best-fit results.
    (b): Same as panel~(a), except now for 104 \emph{Swift}
    GRBs with $z<5$ and $L_{\rm iso}>3.1\times10^{51}$ erg $\rm s^{-1}$.}
\label{fig8}
\end{figure}

With the latest $Swift$ GRBs in hand (see Fig.~\ref{fig4}), we attempt to constrain the minimum halo mass $M_{\rm min}$.
Owing to the influence of the $Swift$ threshold, those low luminosity bursts can not be detected at higher $z$.
To reduce the instrumental selection effect, we also only select luminous bursts with $L_{\rm iso}>1.8\times10^{51}$ erg $\rm s^{-1}$
and $z<4$, as Ref.~\refcite{Kistler et al.2008} did in their analysis. This cut in luminosity and redshift leaves us 118 GRBs,
which are delimited by the red box in Fig.~\ref{fig4}.
Using the $\chi^{2}$ statistic, we can obtain confidence limits in the two-dimensional
($M_{\rm min}$, $\alpha$) plane by fitting the cumulative redshift distribution of these 118 GRBs.
The $1\sigma-3\sigma$ constraint contours of the probability in the
($M_{\rm min}$, $\alpha$) plane are presented in Fig.~\ref{fig8}(a).
These contours show that at the $1\sigma$ level, $-0.54<\alpha<0.99$,
while $M_{\rm min}$ is poorly constrained; only an upper limit of $10^{10.5}$ $\rm M_{\odot}$
can be set at this confidence level.
The cross indicates the best-fit pair $(\log M_{\rm min},\;\alpha)=(7.2,\;-0.15)$.

As shown in Fig.~\ref{fig7}, the SFR $\dot{\rho}_{\star}(z)$ is very sensitive to the
value of $M_{\rm min}$, especially at high-\emph{z}. For the purpose of exploring the
dependence of our results on a possible bias in the high-\emph{z} bursts, we
also consider the sub-sample with $z<5$ and $L_{\rm iso}>L_{\rm lim}(z=5)\approx3.1\times10^{51}$ erg
$\rm s^{-1}$ (consisting of 104 GRBs). These data are delimited by the green
box in Fig.~\ref{fig4}. Compared to the previous sub-sample, this sub-sample has 12
more high-$z$ ($4<z<5$) GRBs. We find that adding 12 high-\emph{z}
GRBs could result in much tighter constraints on $M_{\rm min}$. The contours in Fig.~\ref{fig8}(b) show that
models with $\log M_{\rm min}<7.7$ and $>11.6$ are ruled out at the $1\sigma$ confidence level,
which are in agreement with the results of Ref.~\refcite{Munoz et al.2011}. The evolutionary index is
limited to be $0.10<\alpha<2.55$ ($1\sigma$).
The cross indicates the best-fit pair $(\log M_{\rm min},\;\alpha)=(10.5,\;1.25)$.

In a word, we conclude that only moderate evolution of $(1+z)^{\alpha}$ is consistent with
the observed GRB redshift distribution over $0<z<4$ or $0<z<5$ ($\sim1\sigma$ confidence).
Although the results of us and previous works\cite{Kistler et al.2009} are consistent at the
$1\sigma$ level, we infer a weaker redshift dependence (i.e., weaker enhancement of
the GRB rate compared to the SFR) with lower values of $M_{\rm min}$.
Furthermore, the comparison between Fig.~\ref{fig8}(a) and Fig.~\ref{fig8}(b) can also be found that
the best-fit pairs are very different for the two sub-samples. Because of the increased number of high-\emph{z}
bursts at $4<z<5$, the sub-sample with $L_{\rm iso}>3.1\times10^{51}$ erg $\rm s^{-1}$ and $z<5$ requires a relatively stronger
redshift dependence and a higher value of $M_{\rm min}$.
 Of course, there is still a lot of uncertainty owing to the small high-\emph{z} GRB sample effect.

\section{Conclusions and Discussion}
In this paper, we have briefly reviewed the status for the exploration of the early Universe with GRBs.
A few solid conclusions and discussion can be summarized:

(i) GRBs can be used as standard candles in constructing the Hubble diagram at high-$z$ beyond the
current reach of SNe Ia observations.

However, the dispersion of luminosity relations are still large, which restricted the precision of distance determination
with GRBs. Since the classifications of GRBs are not fully understood, some contamination of the GRB sample in a correlation is
unavoidable, which would make the correlation be dispersed. Note that among all SNe, only a small class SNe (SNe Ia) can be used
as standard candles. On the other hand, the large dispersion may also due to that we
have not yet identified the accurate spectral and lightcurve features to use for the luminosity correlations.
In order to improve the precision of distance measurement, we should take
efforts to investigate the classification problem of GRBs, and
search for more precise luminosity relations, especially the relations with certain physical origins.

(ii) GRBs provide an independent and powerful tool to measure the high-$z$ SFR.

The central difficulty of constraining the high-$z$ SFR with GRBs is that one must know the mechanism responsible
for the difference between the GRB rate and the SFR. The current \emph{Swift} GRB observations can in fact be used to
explore the unknown mechanism. However, the results from Swift data have uncertainties because of
the small GRB sample effect. Fortunately, the upcoming GRB missions such as the Sino-French spacebased multiband
astronomical variable objects monitor (\emph{SVOM}) and the proposed \emph{Einstein Probe} (EP), with wide field of view and high sensitivity,
will be able to discover a sufficiently large number of high-$z$ GRBs. With more abundant observational information in the future,
we will have a better understanding of the mechanism for the SFR-GRB rate discrepancy, and we will measure the high-$z$ SFR
very accurately using the GRB data alone.

(iii) GRBs also constitute a new promising tool for probing star formation in dark matter halos.

In order to constrain the minimum dark matter halo mass $M_{\rm min}$, we also have to know the relation between the GRB rate and the SFR.
In addition to the obvious method of increasing the sample size of GRBs with the help of future dedicated missions, we suggest that
a much more severe constraints on $M_{\rm min}$ will be achieved by combining several independent observations, such as the
observational data of star formation history, the luminosity distribution of galaxies, and the redshift distribution of GRBs.

\section*{Acknowledgments}

This work is partially supported by the National Basic Research Program (``973'' Program)
of China (Grants 2014CB845800 and 2013CB834900), the National Natural Science Foundation
of China (grants Nos. 11322328 and 11373068), the Youth Innovation Promotion Association
(2011231), and the Strategic Priority Research Program ``The Emergence of Cosmological
Structures'' (Grant No. XDB09000000) of the Chinese Academy of Sciences.

\end{document}